\documentclass[onecolumn]{article}
\usepackage{amssymb,amsmath,amsthm}
\usepackage{mathrsfs}
\usepackage[plainpages=false]{hyperref}
\usepackage{url}
\usepackage{bbm}   
\usepackage{graphicx}
\usepackage[left=2.5cm,right=2.5cm,top=4cm,bottom=4cm]{geometry}
\usepackage{subfigure}
\usepackage{multirow}
\usepackage{color}

\usepackage[inline,marginclue,silent,final]{fixme}

\usepackage{paralist}

\usepackage[all]{xy}
\input xy
\xyoption{all}
\xyoption{arc}
\xyoption{line}

\newcommand{\beq}{\begin{displaymath}}
\newcommand{\eeq}{\end{displaymath}}
\newcommand{\beqn}{\begin{equation}}
\newcommand{\eeqn}{\end{equation}}
\newcommand{\beqa}{\begin{eqnarray*}}
\newcommand{\eeqa}{\end{eqnarray*}}
\newcommand{\beqna}{\begin{eqnarray}}
\newcommand{\eeqna}{\end{eqnarray}}

\newcommand{\eq}[1]{~(\ref{#1})}

\theoremstyle{definition} 

\title{Quantum analogues of Hardy's nonlocality paradox}
\author{Tobias Fritz\\
\small{ICFO-Institut de Ciencies Fotoniques, Mediterranean Technology Park, 08860 Castelldefels (Barcelona), Spain}\\[.3cm]
\small{\texttt{tobias.fritz@icfo.es}}}

\begin{document}

\maketitle

\begin{abstract}
Hardy's nonlocality is a ``nonlocality proof without inequalities'': it exemplifies that quantum correlations can be qualitatively stronger than classical correlations. This paper introduces variants of Hardy's nonlocality in the CHSH scenario which are realized by the PR-box, but not by quantum correlations. Hence this new kind of Hardy-type nonlocality is a proof without inequalities showing that superquantum correlations can be qualitatively stronger than quantum correlations.
\end{abstract}

\section{Introduction}

Entanglement and quantum nonlocality are phenomena which have received widespread attention in recent years. Hardy's nonlocality paradox is an especially simple entanglement-witnessing phenomenon; it shows that quantum correlations generated from entangled states can be qualitatively different from correlations generated by local hidden variables. Although it is not a paradox in the strict sense, it is so counterintuitive that a layman unfamiliar with the quantum-mechanical formalism would most likely declare this kind of correlation as outright impossible. 

On the other hand, there are correlations that are consistent with the physical principle of no-signaling, but yet are stronger than all correlations allowed quantum-mechanically. For example, many Bell inequalities detecting the impossibility of local hidden variable models have maximally allowed quantum violations which are smaller than what is compatible with the no-signaling principle. So in general, one has to distinguish three main types of correlations, forming a hierarchy as follows:
\beq
\textrm{local hidden variable correlations}\subset\textrm{quantum correlations}\subset\textrm{no-signaling correlations}
\eeq
Hardy's nonlocality now shows that the first of these two inclusions is qualitatively strict in the sense that some quantum correlations have qualitatively different properties from local hidden variable correlations. Due to this qualitative nature, it has also been called a ``nonlocality proof without inequalities''~\cite{Har}.

In this work, several variants of Hardy's nonlocality paradox are introduced which establish an analogous situation for the second inclusion in this hierarchy. This is achieved by trying to strengthen the ``paradox'' by requiring an additional joint outcome probability to vanish. Surprisingly, this rules out any quantum-mechanical realization, no matter which joint outcome probability is chosen. However, three of these strengthened versions are realized by no-signaling correlations which are ``superquantum'' in the sense that they cannot be realized in a quantum-mechanical world. Hence, there is a ``proof without inequalities'' for the existence of superquantum no-signaling correlations.

This article is structured as follows. After an introductory section on the CHSH scenario and Hardy's nonlocality paradox, the main section on imposing additional vanishing constraints discusses the four conceivable extensions of Hardy's nonlocality and shows that none of them can occur quantum-mechanically, while three of them do occur for the PR-box (Popescu-Rohrlich box~\cite{PopRohr}). Finally, the conclusion provides a short summary and outlines an idea for future research.

\section{Hardy's nonlocality paradox}

The simplest setup that is able to demonstrate quantum nonlocality is the ``CHSH'' scenario conceived by Clauser, Horne, Shimony and Holt~\cite{CHSH}: two parties, commonly dubbed \emph{Alice} and \emph{Bob}, each operate with a physical system of their own on which they respectively conduct one of two dichotomic (i.e. two-valued) measurements. Then quantum theory entails phenomena without classical counterparts: many quantum states with \emph{entanglement} let Alice and Bob observe correlations between their measurements which cannot be explained by classical models defined in terms of local hidden variables; this non-classicality can be detected by observing violations of the \emph{CHSH inequalities}. Alternatively, it can also be demonstrated by observing the occurence of Hardy's nonlocality paradox, although this is more difficult to achieve.

What now follows is a brief mathematical description of the CHSH scenario and the notation used in this work. Alice and Bob can choose between two $\pm 1$-valued observables $A_1, A_2$ and $B_1, B_2$, respectively. The state space of the system under consideration can be of any dimension, and hence the observables may be of any degree of degeneracy. Since we do not want to carry tensor product symbols all around, we use the shorthand notation
\beq
a_k\equiv A_k\otimes\mathbbm{1}\:,\qquad b_l\equiv \mathbbm{1}\otimes B_l\:,\qquad\forall k,l=1,2\:,
\eeq
so that a joint measurement can be written as a product
\beq
A_k\otimes B_l=(A_k\otimes\mathbbm{1})(\mathbbm{1}\otimes B_l)=a_kb_l\:,
\eeq
and the relevant property guaranteeing compatibility of Alice's observables with Bob's observables is the commutativity relation
\beqn
\label{comm}
a_kb_l=b_la_k\qquad\forall k,l=1,2\:.
\eeqn

So consider now the case that Alice decides to measure the observable $a_k$ for $k\in\{1,2\}$, while Bob chooses $b_l$ for $l\in\{1,2\}$. Then the probability of getting a particular outcome $r\in\{-1,+1\}$ for Alice together with a particular outcome $s\in\{-1,+1\}$ for Bob is denoted by
\beq
P(r,s|k,l).
\eeq
For a given shared quantum state $|\psi\rangle$, this probability takes on the form
\beq
P(r,s|k,l)=\left\langle\psi\left|\frac{1+ra_k}{2}\cdot\frac{1+sb_l}{2}\right|\psi\right\rangle
\eeq
since $(1+ra_k)/2$ is the projection operator onto the outcome $r$ for Alice's observable $a_k$, and similar for Bob's observable $b_l$. Hence we have that
\beqn
\label{vanishcond}
P(r,s|k,l)=0\qquad\textrm{if and only if}\qquad(1+ra_k)(1+sb_l)|\psi\rangle=0.
\eeqn

Hardy's nonlocality~(\cite{Har},~\cite{Mer}) occurs when the joint probabilities have the following properties (see fig.~\ref{hardy}):
\begin{align}
P(+1,+1|1,1)=0\label{1}\\
P(-1,-1|1,2)=0\label{2}\\
P(-1,-1|2,1)=0\label{3}\\
P(-1,-1|2,2)>0\label{4}
\end{align}
Of course, it can also occur in many variants which one obtains by permuting the outcome labels of some measurements ($-1\leftrightarrow +1$) or permuting some of the measurements ($a_1\leftrightarrow a_2$, $b_1\leftrightarrow b_2$), or both.

The properties\eq{1}--(\ref{4}) together are impossible in any local hidden variable theory. For suppose that there would exist a local hidden variable model displaying this behavior; it is well-known~\cite{Fine} that this means precisely that these correlations can be described by a statistical combination of the $16$ deterministic models
\beqn
\label{realistic}
a_1^\pm a_2^\pm b_1^\pm b_2^\pm
\eeqn
where each sign stands for the corresponding measurement outcome it determines with certainty, and the four signs can be chosen independently of each other. 

By the assumption\eq{4}, we know that this statistical mixture contains at least one of the four states of the form
\beq
a_1^\pm a_2^- b_1^\pm b_2^-.
\eeq
But now due to\eq{2}, this cannot be one of the two states $a_1^- a_2^- b_1^\pm b_2^-$. Likewise by\eq{3}, it cannot be one of the two states $a_1^\pm a_2^- b_1^- b_2^-$. Therefore, the statistical mixture of deterministic models necessarily contains the model
\beq
a_1^+ a_2^- b_1^+ b_2^-
\eeq
but now this contradicts the assumption\eq{1}! The fact that\eq{1}-(\ref{4}) is impossible for local hidden variable correlations, but is theoretically possible with quantum correlations~\cite{Har},~\cite{Mer} and has been observed in experiment~\cite{Exp}, is therefore referred to as Hardy's nonlocality ``paradox''. Of course, it is not an actual logical paradox in the sense of a mathematical inconsistency. It is rather only a counterintuitive phenomenon which seems paradoxical when one makes the local hidden variable assumption without being aware of doing so.

\begin{figure}
\centering
\begin{tabular}{cc|cc|cc}
\multicolumn{6}{c}{}\\
 \multicolumn{2}{c|}{} & \multicolumn{2}{|c|}{$a_1$} & \multicolumn{2}{|c}{$a_2$}\\
\multicolumn{2}{c|}{} & \multicolumn{1}{|c}{$+1$} & \multicolumn{1}{c|}{$-1$} & \multicolumn{1}{|c}{$+1$} & \multicolumn{1}{c}{$-1$}\\
\hline \multirow{2}{*}{$b_1$} & \multicolumn{1}{c|}{$+1$} & \multicolumn{1}{|c}{$0$} & \multicolumn{1}{c|}{$\cdot$} & \multicolumn{1}{|c}{$\cdot$} & \multicolumn{1}{c}{$\cdot$} \\
 & \multicolumn{1}{c|}{$-1$} & \multicolumn{1}{|c}{$\cdot$} & \multicolumn{1}{c|}{$\cdot$} & \multicolumn{1}{|c}{$\cdot$} & \multicolumn{1}{c}{$0$} \\
\hline \multirow{2}{*}{$b_2$} & \multicolumn{1}{c|}{$+1$} & \multicolumn{1}{|c}{$\cdot$} & \multicolumn{1}{c|}{$\cdot$} & \multicolumn{1}{|c}{$\cdot$} & \multicolumn{1}{c}{$\cdot$} \\
 & \multicolumn{1}{c|}{$-1$} & \multicolumn{1}{|c}{$\cdot$} & \multicolumn{1}{c|}{$0$} & \multicolumn{1}{|c}{$\cdot$} & \multicolumn{1}{c}{$>0$} \\
\end{tabular}
\caption{Conditions on the joint outcome probabilities for Hardy's nonlocality ``paradox'' to occur (up to symmetry). The dotted entries are unspecified and may vanish or not vanish.}
\label{hardy}
\end{figure}

\begin{figure}
\centering
\begin{tabular}{cc|cc|cc}
\multicolumn{6}{c}{}\\
 \multicolumn{2}{c|}{} & \multicolumn{2}{|c|}{$a_1$} & \multicolumn{2}{|c}{$a_2$}\\
\multicolumn{2}{c|}{} & \multicolumn{1}{|c}{$+1$} & \multicolumn{1}{c|}{$-1$} & \multicolumn{1}{|c}{$+1$} & \multicolumn{1}{c}{$-1$}\\
\hline \multirow{2}{*}{$b_1$} & \multicolumn{1}{c|}{$+1$} & \multicolumn{1}{|c}{$0$} & \multicolumn{1}{c|}{$\frac{1}{2}$} & \multicolumn{1}{|c}{$0$} & \multicolumn{1}{c}{$\frac{1}{2}$} \\
 & \multicolumn{1}{c|}{$-1$} & \multicolumn{1}{|c}{$\frac{1}{2}$} & \multicolumn{1}{c|}{$0$} & \multicolumn{1}{|c}{$\frac{1}{2}$} & \multicolumn{1}{c}{$0$} \\
\hline \multirow{2}{*}{$b_2$} & \multicolumn{1}{c|}{$+1$} & \multicolumn{1}{|c}{$0$} & \multicolumn{1}{c|}{$\frac{1}{2}$} & \multicolumn{1}{|c}{$\frac{1}{2}$} & \multicolumn{1}{c}{$0$} \\
 & \multicolumn{1}{c|}{$-1$} & \multicolumn{1}{|c}{$\frac{1}{2}$} & \multicolumn{1}{c|}{$0$} & \multicolumn{1}{|c}{$0$} & \multicolumn{1}{c}{$\frac{1}{2}$} \\
\end{tabular}
\caption{Table of joint outcome probabilities for the PR-box.}
\label{PRbox}
\end{figure}

\section{Imposing additional vanishing constraints}

While the existence of quantum probabilities that satisfy\eq{1}-(\ref{4}) is intriguing, maybe it would be possible to strengthen the underlying ``paradox'' even further by additionally requiring a fourth joint probability $P(r,s|k,l)$ (for some appropriate $r,s,k,l$) to vanish while retaining the existence of a quantum-mechanical model? We will see in the following that this is not the case. But although while none of these stronger ``paradoxes'' actually occurs in quantum theory, several of them do occur for more general nonsignaling boxes, in particular for the PR-box; recall from~\cite{PopRohr} that the joint probabilities for the PR-box are those displayed in fig.~\ref{PRbox}.

Hence we are going to postulate an additional $0$ in the table of joint probabilities fig.~\ref{hardy}. This is not interesting as long as the new $0$ is directly adjacent to an existing $0$ within the corresponding $2\times 2$-subtable: in this case, the no-signaling constraint forces an entire row or column in the full table to vanish, leading to the effect that one of the four observables has a definite outcome, and hence precluding the possibility of observing quantum nonlocality. Figure~\ref{superhardy} shows the remaining cases, up to symmetry, numbered as Hardy$N$ for $N=1,2,3,4$. Each case will now be treated, one at a time, and it will be shown that all of these are impossible to occur in quantum theory, while Hardy1, Hardy2 and Hardy3 do occur for the PR-box.

While proving the quantum impossibility would possibly be easier by using the fact that one can assume each party's subsystem to be a qubit~\cite{Mas}, the proofs presented here are independent of that result.

\begin{figure}
\centering
\subfigure[Hardy1]{
	\label{Hardy1}
\begin{tabular}{cc|cc|cc}
\multicolumn{6}{c}{}\\
 \multicolumn{2}{c|}{} & \multicolumn{2}{|c|}{$a_1$} & \multicolumn{2}{|c}{$a_2$}\\
\multicolumn{2}{c|}{} & \multicolumn{1}{|c}{$+1$} & \multicolumn{1}{c|}{$-1$} & \multicolumn{1}{|c}{$+1$} & \multicolumn{1}{c}{$-1$}\\
\hline \multirow{2}{*}{$b_1$} & \multicolumn{1}{c|}{$+1$} & \multicolumn{1}{|c}{$0$} & \multicolumn{1}{c|}{$\cdot$} & \multicolumn{1}{|c}{$\cdot$} & \multicolumn{1}{c}{$\cdot$} \\
 & \multicolumn{1}{c|}{$-1$} & \multicolumn{1}{|c}{$\cdot$} & \multicolumn{1}{c|}{$\textcolor{blue}0$} & \multicolumn{1}{|c}{$\cdot$} & \multicolumn{1}{c}{$0$} \\
\hline \multirow{2}{*}{$b_2$} & \multicolumn{1}{c|}{$+1$} & \multicolumn{1}{|c}{$\cdot$} & \multicolumn{1}{c|}{$\cdot$} & \multicolumn{1}{|c}{$\cdot$} & \multicolumn{1}{c}{$\cdot$} \\
 & \multicolumn{1}{c|}{$-1$} & \multicolumn{1}{|c}{$\cdot$} & \multicolumn{1}{c|}{$0$} & \multicolumn{1}{|c}{$\cdot$} & \multicolumn{1}{c}{$>0$} \\\\
	\end{tabular}\hspace{2cm}}
\subfigure[Hardy2]{
	\label{Hardy2}
\begin{tabular}{cc|cc|cc}
\multicolumn{6}{c}{}\\
 \multicolumn{2}{c|}{} & \multicolumn{2}{|c|}{$a_1$} & \multicolumn{2}{|c}{$a_2$}\\
\multicolumn{2}{c|}{} & \multicolumn{1}{|c}{$+1$} & \multicolumn{1}{c|}{$-1$} & \multicolumn{1}{|c}{$+1$} & \multicolumn{1}{c}{$-1$}\\
\hline \multirow{2}{*}{$b_1$} & \multicolumn{1}{c|}{$+1$} & \multicolumn{1}{|c}{$0$} & \multicolumn{1}{c|}{$\cdot$} & \multicolumn{1}{|c}{$\cdot$} & \multicolumn{1}{c}{$\cdot$} \\
 & \multicolumn{1}{c|}{$-1$} & \multicolumn{1}{|c}{$\cdot$} & \multicolumn{1}{c|}{$\cdot$} & \multicolumn{1}{|c}{$\cdot$} & \multicolumn{1}{c}{$0$} \\
\hline \multirow{2}{*}{$b_2$} & \multicolumn{1}{c|}{$+1$} & \multicolumn{1}{|c}{$\textcolor{blue}0$} & \multicolumn{1}{c|}{$\cdot$} & \multicolumn{1}{|c}{$\cdot$} & \multicolumn{1}{c}{$\cdot$} \\
 & \multicolumn{1}{c|}{$-1$} & \multicolumn{1}{|c}{$\cdot$} & \multicolumn{1}{c|}{$0$} & \multicolumn{1}{|c}{$\cdot$} & \multicolumn{1}{c}{$>0$} \\\\
	\end{tabular}\hspace{2cm}\vspace{1cm}}
\subfigure[Hardy3]{
	\label{Hardy3}
\begin{tabular}{cc|cc|cc}
\multicolumn{6}{c}{}\\
 \multicolumn{2}{c|}{} & \multicolumn{2}{|c|}{$a_1$} & \multicolumn{2}{|c}{$a_2$}\\
\multicolumn{2}{c|}{} & \multicolumn{1}{|c}{$+1$} & \multicolumn{1}{c|}{$-1$} & \multicolumn{1}{|c}{$+1$} & \multicolumn{1}{c}{$-1$}\\
\hline \multirow{2}{*}{$b_1$} & \multicolumn{1}{c|}{$+1$} & \multicolumn{1}{|c}{$0$} & \multicolumn{1}{c|}{$\cdot$} & \multicolumn{1}{|c}{$\cdot$} & \multicolumn{1}{c}{$\cdot$} \\
 & \multicolumn{1}{c|}{$-1$} & \multicolumn{1}{|c}{$\cdot$} & \multicolumn{1}{c|}{$\cdot$} & \multicolumn{1}{|c}{$\cdot$} & \multicolumn{1}{c}{$0$} \\
\hline \multirow{2}{*}{$b_2$} & \multicolumn{1}{c|}{$+1$} & \multicolumn{1}{|c}{$\cdot$} & \multicolumn{1}{c|}{$\cdot$} & \multicolumn{1}{|c}{$\cdot$} & \multicolumn{1}{c}{$\cdot$} \\
 & \multicolumn{1}{c|}{$-1$} & \multicolumn{1}{|c}{$\cdot$} & \multicolumn{1}{c|}{$0$} & \multicolumn{1}{|c}{$\textcolor{blue}0$} & \multicolumn{1}{c}{$>0$} \\\\
	\end{tabular}\hspace{2cm}}
\subfigure[Hardy4]{
	\label{Hardy4}
\begin{tabular}{cc|cc|cc}
\multicolumn{6}{c}{}\\
 \multicolumn{2}{c|}{} & \multicolumn{2}{|c|}{$a_1$} & \multicolumn{2}{|c}{$a_2$}\\
\multicolumn{2}{c|}{} & \multicolumn{1}{|c}{$+1$} & \multicolumn{1}{c|}{$-1$} & \multicolumn{1}{|c}{$+1$} & \multicolumn{1}{c}{$-1$}\\
\hline \multirow{2}{*}{$b_1$} & \multicolumn{1}{c|}{$+1$} & \multicolumn{1}{|c}{$0$} & \multicolumn{1}{c|}{$\cdot$} & \multicolumn{1}{|c}{$\cdot$} & \multicolumn{1}{c}{$\cdot$} \\
 & \multicolumn{1}{c|}{$-1$} & \multicolumn{1}{|c}{$\cdot$} & \multicolumn{1}{c|}{$\cdot$} & \multicolumn{1}{|c}{$\cdot$} & \multicolumn{1}{c}{$0$} \\
\hline \multirow{2}{*}{$b_2$} & \multicolumn{1}{c|}{$+1$} & \multicolumn{1}{|c}{$\cdot$} & \multicolumn{1}{c|}{$\cdot$} & \multicolumn{1}{|c}{$\textcolor{blue}0$} & \multicolumn{1}{c}{$\cdot$} \\
 & \multicolumn{1}{c|}{$-1$} & \multicolumn{1}{|c}{$\cdot$} & \multicolumn{1}{c|}{$0$} & \multicolumn{1}{|c}{$\cdot$} & \multicolumn{1}{c}{$>0$} \\\\
	\end{tabular}\hspace{2cm}}
\caption{The four quantum analogues of Hardy's nonlocality paradox (up to symmetry).}
\label{superhardy}
\end{figure}

\paragraph{Hardy1 (fig.~\ref{Hardy1}).}

So in addition to\eq{1}--(\ref{4}), we now also require that $\color{blue}P(-1,-1|1,1)=0$. By\eq{vanishcond}, these five equations translate into conditions on the quantum state and the quantum observables,
\begin{align}
\label{11}(1+b_1)&(1+a_1)|\psi\rangle=0\\
\label{13}(1-b_1)&(1-a_2)|\psi\rangle=0\\
\label{14}(1-b_2)&(1-a_1)|\psi\rangle=0\\
\label{12}\color{blue}(1-b_1)&\color{blue}(1-a_1)|\psi\rangle=0\\
\label{15}(1-b_2)&(1-a_2)|\psi\rangle\neq 0
\end{align}

In the following calculations, we will several times make implicit use of the commutativity rule\eq{comm}. To start, it follows by adding\eq{11} and\eq{12} that $b_1a_1|\psi\rangle=-|\psi\rangle$, so that $b_1^2=1$ implies that
\beq
b_1|\psi\rangle=-a_1|\psi\rangle
\eeq
With this, equations\eq{13} and\eq{14} are equivalent to, respectively,
\begin{align}
\label{replace1}
(1-a_2)a_1|\psi\rangle=-(1-a_2)|\psi\rangle\\
\label{replace2}
(1-b_2)b_1|\psi\rangle=-(1-b_2)|\psi\rangle
\end{align}
But now it follows that
\begin{align*}
(1-b_2)(1-a_2)|\psi\rangle&=-(1-b_2)(1-a_2)b_1a_1|\psi\rangle\stackrel{(\ref{replace1})}{=}(1-a_2)(1-b_2)b_1|\psi\rangle\\
&\stackrel{(\ref{replace2})}{=}-(1-b_2)(1-a_2)|\psi\rangle
\end{align*}
so that comparing the left-hand side with the right-hand side shows that $(1-b_2)(1-a_2)|\psi\rangle=0$, in contradiction to\eq{15}. Therefore, the unintuitive situation Hardy1 is impossible in quantum theory. However, the PR box fig.~\ref{PRbox} does allow this situation to occur: there, the value of the ``paradoxical'' probability $P(-1,-1|2,2)$ is even as big as $\tfrac{1}{2}$. So even if the situation Hardy1 is not quantum, it is clearly compatible with the no-signaling principle.

\paragraph{Hardy2 (fig.~\ref{Hardy2}).}

We now postulate $\color{blue}P(+1,+1|1,2)=0$ in addition to\eq{1}--(\ref{4}). Hence we obtain the following requirements on the quantum state and the quantum observables,
\begin{align}
\label{21}(1+b_1)&(1+a_1)|\psi\rangle=0\\
\label{22}(1-b_1)&(1-a_2)|\psi\rangle=0\\
\label{23}(1-b_2)&(1-a_1)|\psi\rangle=0\\
\label{24}\color{blue}(1+b_2)&\color{blue}(1+a_1)|\psi\rangle=0\\
\label{25}(1-b_2)&(1-a_2)|\psi\rangle\neq 0
\end{align}
By adding\eq{23} and\eq{24}, we can observe the perfect anticorrelation
\beq
b_2|\psi\rangle=-a_1|\psi\rangle.
\eeq
Then we get for the desired outcome probability $P(-1,-1|2,2)$,
\begin{align*}
\langle\psi|(1-b_2)(1-a_2)|\psi\rangle=&\phantom{-}\:\:\langle\psi|(1+a_1)(1-a_2)|\psi\rangle\stackrel{(\ref{21}),(\ref{22})}{=}-\langle\psi|(1+a_1)b_1b_1(1-a_2)|\psi\rangle\\
=&-\langle\psi|(1+a_1)(1-a_2)|\psi\rangle=0.
\end{align*}
So also this strengthening of premises does not yield a quantum-mechanical ``paradox'', although the PR-box fig.~\ref{PRbox} shows this behavior.

\paragraph{Hardy3 (fig.~\ref{Hardy3}).}

In this case we also want $\color{blue}P(+1,-1|2,2)=0$, so that we ought to start from
\begin{align}
\label{31}(1+b_1)&(1+a_1)|\psi\rangle=0\\
\label{32}(1-b_1)&(1-a_2)|\psi\rangle=0\\
\label{33}(1-b_2)&(1-a_1)|\psi\rangle=0\\
\label{34}\color{blue}(1-b_2)&\color{blue}(1+a_2)|\psi\rangle=0\\
\label{35}(1-b_2)&(1-a_2)|\psi\rangle\neq 0
\end{align}
which gives upon inserting the decompositions $1=\tfrac{1}{2}(1-a_j)+\tfrac{1}{2}(1+a_j)$,
\begin{align*}
\langle\psi|(1-b_1)(1-b_2)|\psi\rangle=\tfrac{1}{2}\langle\psi|(1-b_1)(1-a_2)(1-b_2)|\psi\rangle+\tfrac{1}{2}\langle\psi|(1-b_1)(1+a_2)(1-b_2)|\psi\rangle\stackrel{(\ref{32}),(\ref{34})}=0+0=0\\
\langle\psi|(1+b_1)(1-b_2)|\psi\rangle=\tfrac{1}{2}\langle\psi|(1+b_1)(1-a_1)(1-b_2)|\psi\rangle+\tfrac{1}{2}\langle\psi|(1+b_1)(1+a_1)(1-b_2)|\psi\rangle\stackrel{(\ref{33}),(\ref{31})}=0+0=0
\end{align*}
But the quantity we are interested in is given in terms of
\begin{align*}
\langle\psi|(1-a_2)(1-b_2)|\psi\rangle&\stackrel{(\ref{24})}{=}2\langle\psi|(1-b_2)|\psi\rangle\\
&\:\:=\:\langle\psi|(1-b_1)(1-b_2)|\psi\rangle+\langle\psi|(1+b_1)(1-b_2)|\psi\rangle=0+0=0
\end{align*}
Thus also in this case, quantum-mechanical nonlocality cannot achieve such correlations, although the PR-box fig.~\ref{PRbox} does so.

\paragraph{Hardy4 (fig.~\ref{Hardy4}).}
On first look, these conditions on the table of joint probabilities seem to be compatible with the no-signaling principle. However, a direct check by inspection shows that neither the PR-box from fig.~\ref{PRbox} nor any of its seven siblings (switch $a_1\leftrightarrow a_2$ and/or $b_1\leftrightarrow b_2$ and/or $-1\leftrightarrow +1$ for Bob) display this kind of nonlocality. We now claim that the same holds for any statistical mixture of PR-boxes and/or local hidden variable models, and will prove this claim below. This then shows that no no-signaling theory can display Hardy4, since the vertices of the no-signaling polytope are precisely the deterministic local hidden variable models together with the PR-boxes~\cite{Blal}. In particular, again no quantum paradoxes are possible.

In order to prove the claim, it is enough to check that if two nonsignaling boxes do not display Hardy4, then neither does any statistical combination of these two. But this is simple: if one of the boxes does not display Hardy4 because it violates one of the vanishing constraints by having a positive probability for the corresponding outcome, then this will also hold for any non-trivial statistical combination of the two boxes. So we are left with the case that both boxes satisfy all the vanishing constraints. Since by assumption these boxes do not display Hardy4, they need to have the property that also the paradoxical probability vanishes, $P(-1,-1|2,2)=0$. But then, this also holds for any statistical combination of the two boxes.

\paragraph{Characterizing the PR-box.} Another question to ask is how many additional zeroes one would have to add to Hardy's nonlocality fig.~\ref{hardy} so that the PR-box becomes the only set of correlations showing this behavior. Despite quantum-mechanical impossibility, a set of correlations having the property Hardy$N$ for some $N\in\{1,2,3\}$ can still be far from being a PR-box. This can be seen for example in the following way: recall that a deterministic local hidden variable model is defined by specifying a sign for each of the four observables $a_1,a_2,b_1,b_2$. In each $2\times 2$-subtable, such an assignment selects one of the four possible outcomes as the definite one. Now for Hardy$N$, we can take the PR-box fig.~\ref{PRbox} and form a statistical mixture with a deterministic local hidden variable theory that satisfies all vanishing constraints, but not the positivity condition $P(-1,-1|2,2)>0$. This statistical mixture still has the property Hardy$N$, but is different from the PR-box. Now it is not hard to see by inspection of fig.~\ref{superhardy} that for each $N\in\{1,2,3\}$, there are many choices of deterministic local hidden variable models which are consistent with the vanishing probabiliti/s required for Hardy$N$. Therefore, really characterizing the PR-box or a statistical mixture of several PR-boxes without a contribution of deterministic local hidden variable models requires the introduction of several additional zeroes in Hardy$N$ for each $N$.

\section{Conclusion}

The present results manifest that quantum nonlocality is also qualitatively of a very special kind. Although it is possible for unintuitive ``paradoxes'' like Hardy's nonlocality to occur, one can formulate stronger versions of these ``paradoxes'' which do not occur in quantum theory, although they do occur in theories with superquantum correlations.

As a possible direction for future research\footnote{due to Antonio Ac\'in.}, one might try to look at the extensions of Hardy's nonlocality introduced in~\cite{BBMH}. There, the authors have found variants of Hardy's paradox which use $M$ dichotomic observables per site instead of only $2$. The contradiction with local hidden variable models arises upon requiring that $2M-1$ appropriate joint outcome probabilities vanish, while one joint outcome probability needs to be strictly positive. So now how many how many additional zeroes can be introduced and in which positions without relegating the existence of a quantum-mechanical model? Saying anything about this seems much harder than for the CHSH scenario. Possibly a closer analysis of the proofs presented here might reveal general techniques for answering questions like this.

\paragraph{Acknowledgements.} The author would like to thank Sibasish Ghosh for introducing him to Hardy's nonlocality paradox. Antonio Ac\'in has kindly made several important suggestions on how to improve an earlier version of this manuscript and pointed to~\cite{BBMH}. Two anonymous referees have also made helpful comments regarding the presentation of the results. This work was conducted while the author was a member of the IMPRS graduate program at the Max Planck Institute for Mathematics.

\bibliographystyle{halpha.bst}
\bibliography{hardys_paradox}

\end{document}